# On the foreshocks of strong earthquakes in the light of catastrophe theory

**A. V. Guglielmi[1], A. L. Sobisevich[1], L. E. Sobisevich[1], and I. P. Lavrov[2],**

[1] *Schmidt Institute of Physics of the Earth, Russian Academy of Sciences, Moscow, Russia*

[2] *Borok Geophysical Observatory, Schmidt Institute of Physics of the Earth, Russian Academy of Sciences, Borok, Yaroslavl oblast, Russia*

E-mail: guglielmi@mail.ru, alex@ifz.ru, sobis@ifz.ru, lavrvania@mail.ru

## Abstract

It is useful to consider the earthquakes in terms of catastrophe theory. In the paper, we illustrate this statement by analysis foreshocks preceding the strong earthquakes. We focused on the so-called catastrophe flags, and on the triggers that cause the critical transitions. The analysis shows a sharp increase of foreshock activity 3 h before the mainshock. This result is reminiscent of the well-known activation of the ULF magnetic precursors of earthquakes. Furthermore, we found that the characteristic frequency of foreshock sequence decreases a few hours before the mainshock, which is consistent with a prediction of the catastrophe theory. Finally, the analysis testifies that the strong foreshocks seems may be triggers of mainshocks. The idea is that the surface waves propagating outwards from the foreshock return back to the vicinity of the epicenter after having made a complete revolution around the Earth and induce there the mainshock.

Contents

## 1. Introduction

The concepts and ideas of the catastrophe theory are useful tools for analyzing earthquakes. This theory has been created in sixties of the last century (e.g., see [Thom, 1976; Zeeman, 1976; Gilmore, 1981; Horsthemke, Lefever, 1984; Arnold, 1992]). We recall that the theory studies the singularities of smooth maps and bifurcations of dynamical systems. State of the system is described by a set of functions $\psi_j(t, c_\alpha)$, $j = \overline{1,n}$, which depend on the time and the so-called control parameters $c_\alpha$, $\alpha = \overline{1,m}$. The central role in the catastrophe theory is assigned to the potential function $U(\psi_j, c_\alpha)$ which depends on the state and the control parameters. The evolution of system is described by the equations

$$\frac{d\psi_j}{dt} = -\frac{\partial U}{\partial \psi_j}. \qquad (1)$$

The most important task is to analyze the critical points of the potential function. These are the points at which not only the first derivatives, but also higher-order derivatives vanish. In particular, the Hessian of the potential function vanishes at the critical point:

$$\det\left(\frac{\partial^2 U}{\partial \psi_i \partial \psi_j}\right) \to 0\,. \qquad (2)$$

The catastrophe comes as a sharp qualitative change of the state of dynamical system in the course a continuous quantitative variation of the control parameters. The theory predicts a number of signs of the approaching disaster. These signs (or symptoms) are called “catastrophe flags”, bearing in mind that the dynamical system as if “warns” concerning the impending threat [Gilmore, 1981]. The first task of our work is to find the “flags” in the flow of foreshocks during the last few hours before a strong earthquake. This problem is studied in the Section 2 and Section 3. The second task of our work relates to the field of study of the triggers earthquakes. In the Section 4 we are considering the endogenous and exogenous triggers in the frame of simple catastrophe model, and in the Section 5 we are analyzing the round-the-world seismic echo as a possible trigger of the mainshock.

Initial data for the study were taken from the earthquake catalog of the International Seismological Centre (ISC, http://www.isc.ac.uk). We used the method of superposed epoch [Samson, Yeung, 1986] for the pilot analysis of foreshocks from 1964 to 2009. In addition, we have also used the method of spectral-temporal analysis for revealing of the phenomenon of mode softening before the onset of critical transition.

## 2. Activization of foreshocks before earthquake

Recall that the foreshocks as the mechanical precursors of strong earthquakes occur in the final stage of seismic cycle [Sobolev et al., 2009]. At the same stage there are a variety of the electromagnetic precursors of coming earthquake (e.g., see [Fraser-Smith et al., 1990; Hattori, 2004; Pulinets, 2004; Sobisevich, Sobisevich, 2010; Sobisevich et al., 2010; Guglielmi, Zotov, 2012; Hayakawa, 2012; Zotov et al., 2013]). In the context of this work, the observations indicating a sharp activization of ULF magnetic pulsations that occurs in the time interval 2 – 4 h before the strong earthquakes [Sobisevich, Sobisevich, 2010] deserve a special attention. Exactly these observations stimulated our research. We have set the task to find the specific properties of foreshock dynamics which are associated with this activation of ULF pulsations in one way or another. And this still managed to do. Looking ahead, we say that our analysis shows a sharp increase of foreshock activity 3 h before the mainshock.

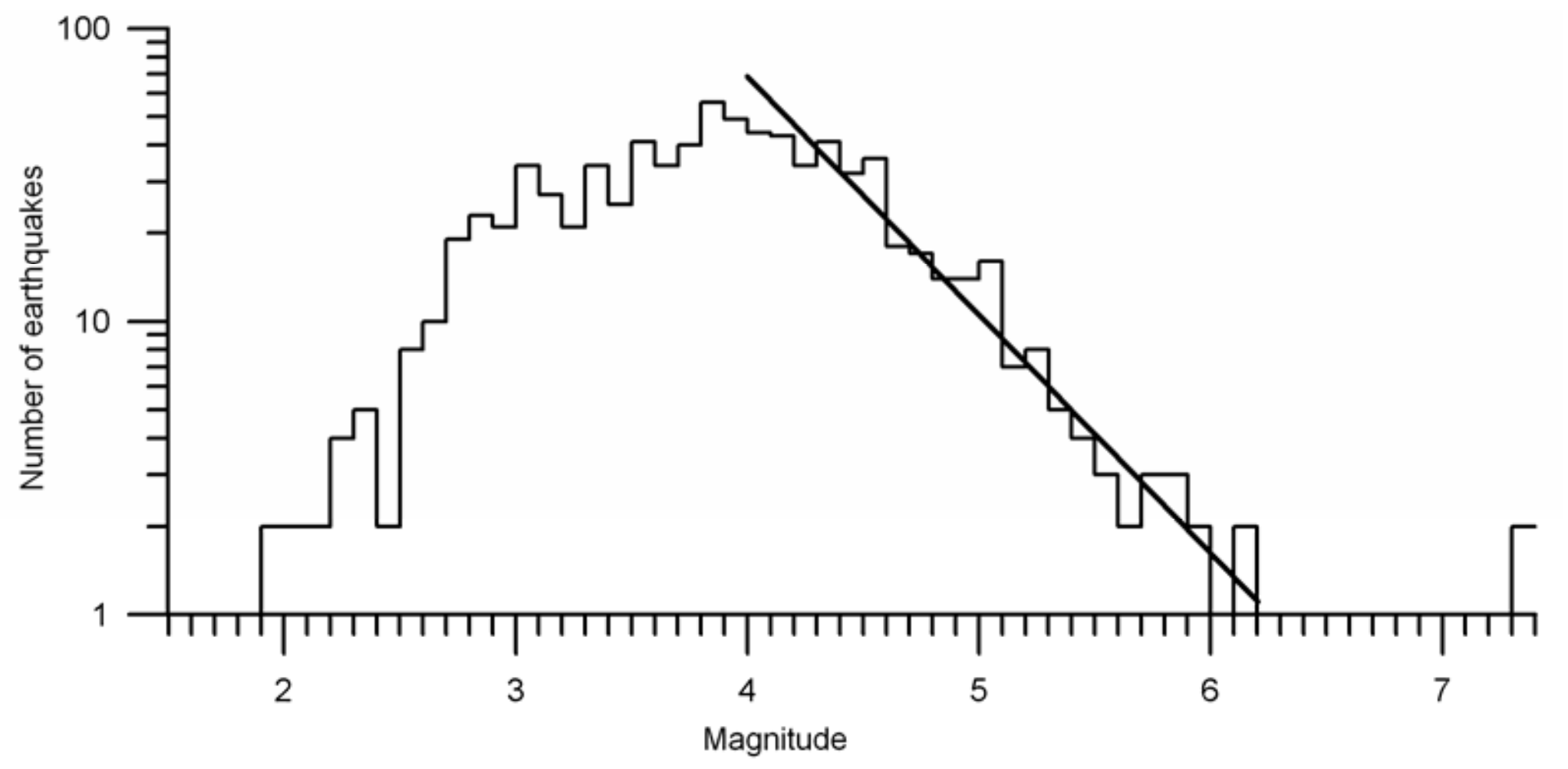

Fig. 1. Magnitude distribution of earthquakes over the intervals 10 h before strong earthquakes ($M \geq 7.5$) in the epicentral zones of radius of 600 km. The straight line approximates the complete part of the catalogue ISC (1964 - 2009).

First of all, the foreshocks were selected in the epicentral zones of the strong mainshocks. By definition, we call mainshock strong if its magnitude $M \geq 7.5$. Fig. 1 shows the magnitude distribution of the foreshocks over the intervals 10 h before mainshocks. The straight line $\lg N = 5.09 - 0.8\ M$ approximates the complete part of the catalogue that corresponds to the magnitudes $M > 3.8$. Here N is the number of foreshocks with a given M. The correlation coefficient between N and M equals to 0.97 in the complete part of distribution.

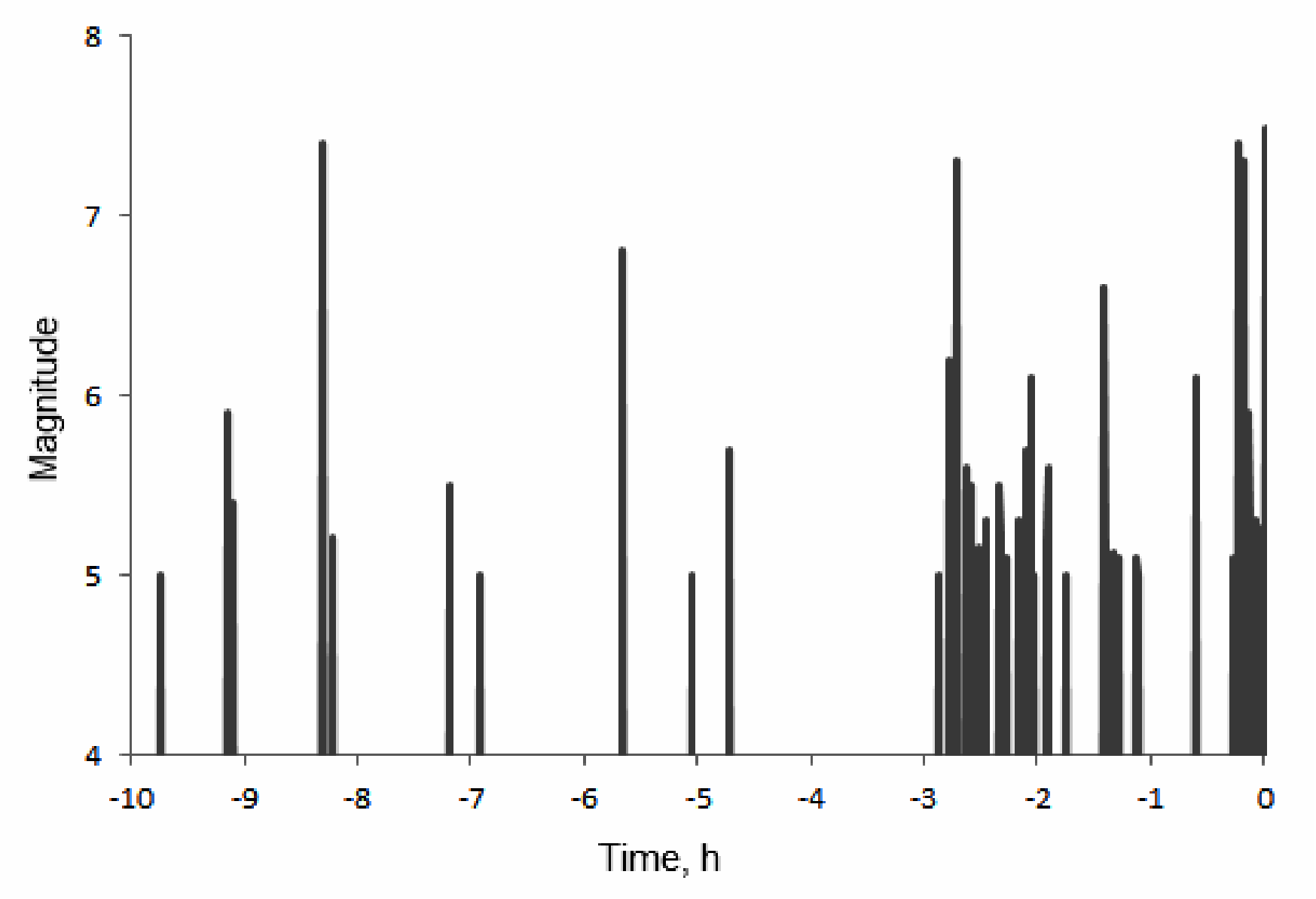

Fig. 2. The strong foreshocks ($5 \le M < 7.5$) in the final stages of preparation of strong earthquakes ($M \ge 7.5$). The zero time stamp corresponds to the mainshock.

Fig. 2 shows the sequence of foreshocks. It was derived by the method of superposed epoch. The moments of mainshocks are chosen as the bench mark for synchronization of the various foreshock sequences. The total number of bench marks is 92. Here we chose only strong foreshocks ($5 \le M < 7.5$) from the complete part of the ISC catalogue (see Fig. 1).

Analyzing Fig. 2, we see that the distribution of foreshocks in time is very uneven. Namely, the sharp upsurge foreshock activity occurs after -3 h. This reminds activation of the ULF magnetic pulsations in the time interval 2 – 4 h before the strong earthquakes [Sobisevich, Sobisevich, 2010]. We are not talking here about the cause-and-effect relationship, as there is currently not sufficient reason for this. We just assert that there is some parallelism of mechanical and electromagnetic processes in the course of preparation of a strong earthquake. In our opinion, this observation sheds a new light on the complex problem of short-term precursors.

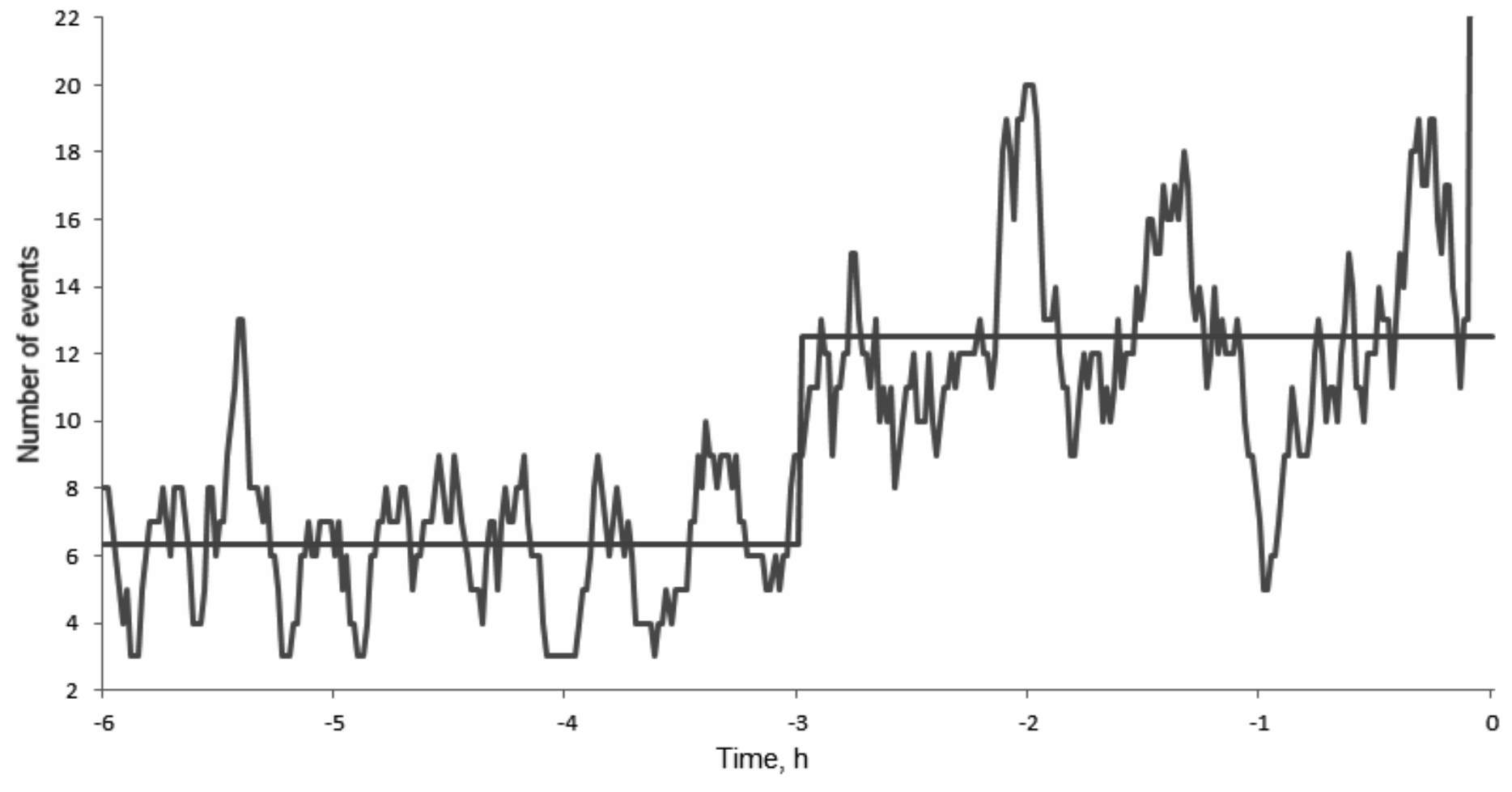

Fig. 3. Averaged dynamics of foreshocks (M < 7.5) in the epicentral zones of the strong earthquakes (M ≥ 7.5). The zero mark corresponds to the mainshock. The horizontal lines represent mean values of foreshocks.

The jump in foreshock activity will occur even if we remove the lower limit on the magnitude of foreshocks. This can be seen in Fig. 3 which gives an idea of the average dynamics of foreshocks with M < 7.5. To construct the figure we calculated the number of foreshocks in the window of width 11 min, which shifts along the horizontal axis by steps of 1 min. The mean values of foreshocks equal 6.3 ± 0.43 until -3 h, and 12.5 ± 0.68 after -3 h. In addition to the jump we see in Fig. 3 the increase of fluctuations with time. This is reminiscent of the critical opalescence which occurs in a phase transition. In the catastrophe theory, this corresponds to the anomalous dispersion near the critical point, or reduction in localization, or deconfinement [Gilmore, 1981].

## 3. Softening of spectrum of the foreshock sequence

It is assumed that the smooth evolution of a dynamical system (a hotbed of impending earthquake) brings it to the threshold, followed by a qualitative jump which called a catastrophe (in our case, it is the formation of the main rupture of rocks). Experience has shown that catastrophes occur often unexpectedly. In practice the prediction of real catastrophe is not possible in many cases. Under such circumstances it is important to know that there is the theory, which is indicating to some signs of approaching a sharp change in the state of a dynamic system.

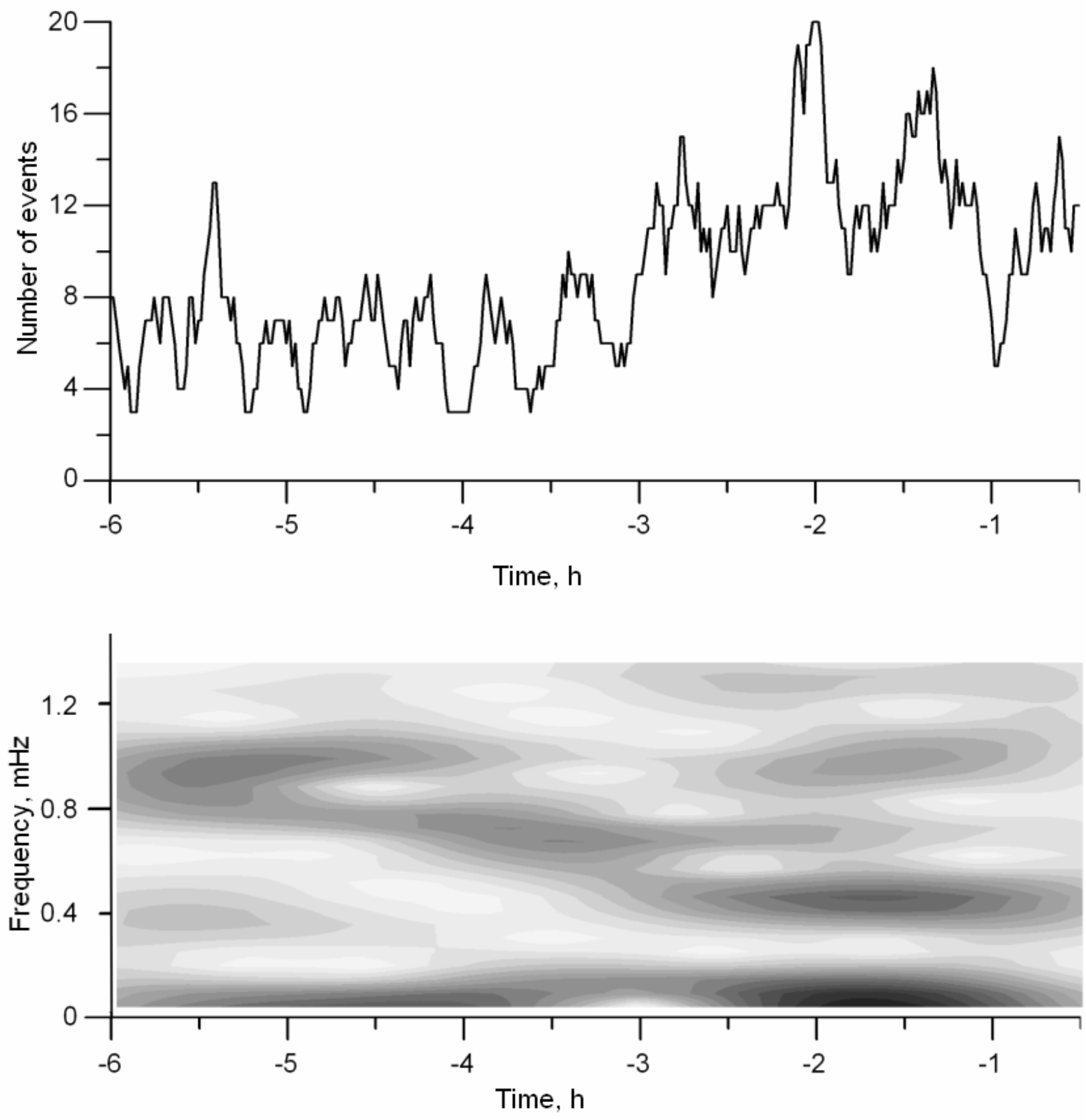

Fig. 4. The variation of the number of foreshocks (upper panel) and the corresponding dynamical spectrum (lower panel).

We chose one of the signs of this kind and tried to find it in the flow of foreshocks. We would like to find the so-called critical slowing down, or mode softening before the onset of the disaster. This is a fairly common feature of the dynamical systems. The point here is as follows. According to (2) the Hessian of the potential function tends to zero at the critical point. Hence it follows that on approaching to a catastrophe at least one of the eigen frequencies of the dynamical system is reduced [Gilmore, 1981]. Apparently, we were able to detect the mode softening before the mainshock. The upper panel of Fig. 4 shows the variation of foreshocks, and the lower shows the corresponding dynamical spectrum. It is reasonable to assume that the dynamic spectrum demonstrates the mode softening near the "rupture point" in accordance with the catastrophe theory.

## 4. Endogenous and exogenous triggers

Following Guglielmi et al. (2013), we consider the endogenous and exogenous triggers in the frame of the fold catastrophe according to Thom’s list of elementary catastrophes [Thom, 1976], or $A_2$ catastrophe according to Arnold’s list [Arnold, 1992]. The potential function is $U(\psi) = U(0) - c_1\psi + c_2\psi^2 - c_3\psi^3$ with $\psi \geq 0$, $c_i \geq 0$, $i = 1, 2, 3$. In other words, the potential has the form of a cubic parabola.

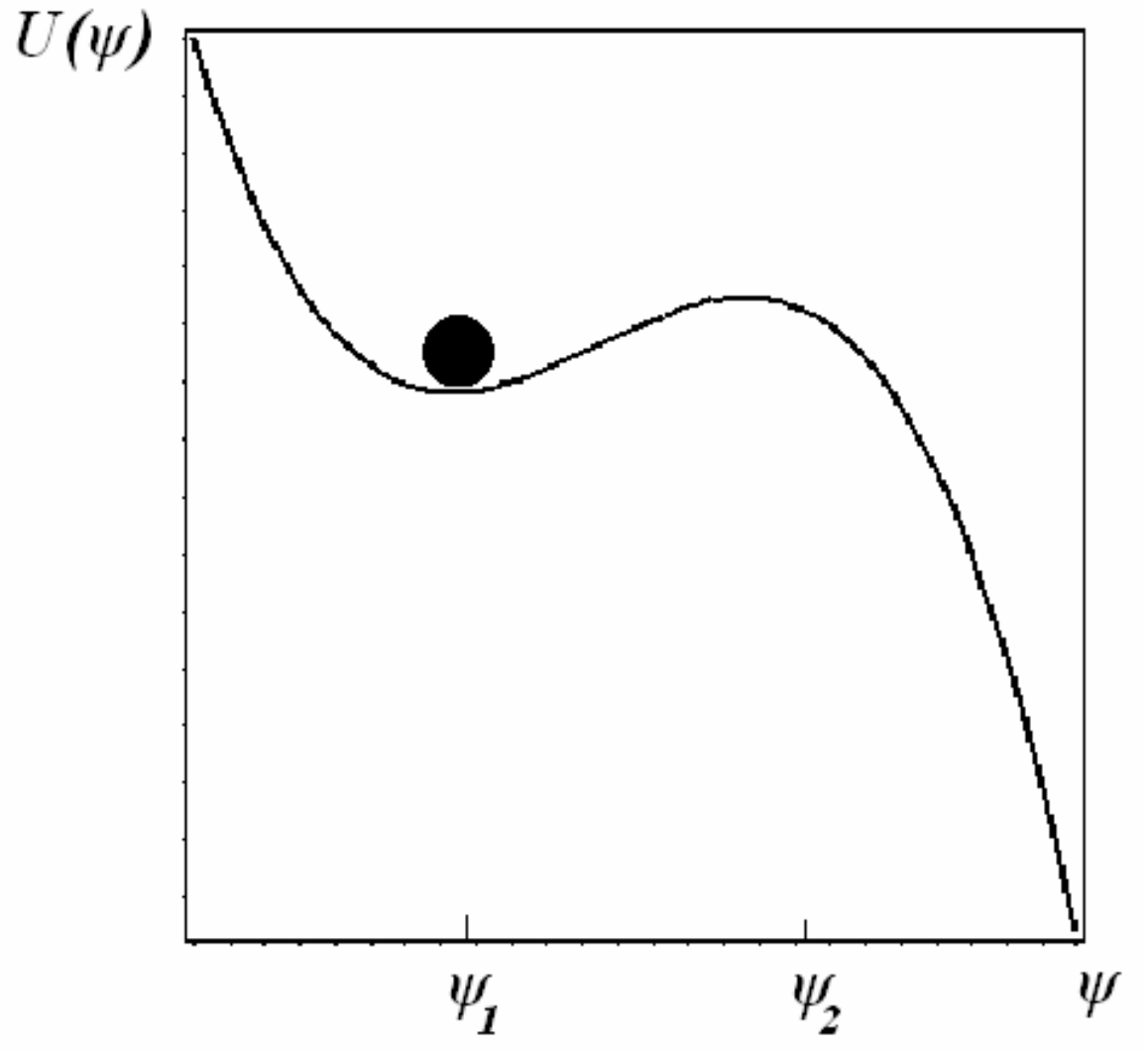

Fig. 5. The typical shape of potential relief. The black ball is in metastable state. The critical transition can be initiated by either endogenous or exogenous trigger.

The minimum and maximum of the potential in Fig. 5 correspond to the stable ($\psi_1$) and unstable ($\psi_2$) equilibrium states ($d\psi / dt = 0$). State $\psi_2$ is metastable. This means that being affected by noise, which always exists in real conditions, the system can make a phase transition $\psi_1 \to \psi > \psi_2$ and thereby lose equilibrium ($d\psi / dt > 0$). The background seismic noise always exists in the Earth's crust or, more exactly, in the future source of the impending earthquake. Under the action of this factor, the earthquake can occur well before the bifurcation, which is determined by the condition $\Delta U = 0$ in the frame of the fold model. Here $\Delta U = U(\psi_2) - U(\psi_1)$ is the potential barrier. Hence, the fluctuation of the stress field of the rocks can become a trigger that initiates the earthquake. Such triggers are naturally referred to as endogenous. The critical transition under the action of an endogenous trigger is called spontaneous transition. In order to describe them, we should use the stochastic Langevin equation

$$\frac{d\psi}{dt} = -\frac{\partial U}{\partial \psi} + \xi(t) \tag{3}$$

instead of the dynamic equation (1). Here, the additive term $\xi(t)$ is a zero-mean random function with $\langle \xi(t')\xi(t'') \rangle = 2D\delta(t' - t'')$, where $\delta(t)$ is the Dirac delta, and the angular brackets denote statistical average. The new phenomenological parameter of the model $D$ is proportional to the intensity of seismic noise in the maturing source.

The further elaboration of the model implies considering the external forces $f(t)$ acting upon the system:

$$\frac{d\psi}{dt} = -\frac{\partial U}{\partial \psi} + \xi(t) + f(t). \quad (4)$$

The critical transition under the action of an external force is natural to refer to as induced, and the corresponding trigger $f(t)$ as exogenic. If $f = 0$, then the probability of the transition is proportional to $\exp(-\Delta U / D)$, i.e. it is exponentially small with a sufficiently high potential barrier [Kramers, 1940]. We note that at $f \neq 0$, the probability of the transition can sharply increase even in the case of relatively small amplitude of external impact. Model (4) indicates that the probability of the transition increases by a factor of $\exp[(F/D)(\psi_2 - \psi_1)]$. Here $F = < 2 f^2 >^{1/2}$ (see [Smelyanskiy et al., 1999], where the important particular case of sine impact is analyzed).

## 5. Round-the-world seismic echo as a trigger of mainshock

The idea of round-the-world seismic echo as an exogenous trigger emerged in the course of analysis of the Sumatra-Andaman earthquake. As is well known, the earthquake with the magnitude M = 9 occurred in the Southeast Asia on December 26, 2004 (e.g., see [Zavyalov, 2005]). It was noticed that the time delay of the strongest aftershock (with the magnitude M = 7.2) is approximately equal to the travel time of the surface wave around the world (it is about 3 h). This suggested that the round-trip seismic echo of the mainshock could have been the trigger that initiated this strong aftershock [Guglielmi et al., 2013].

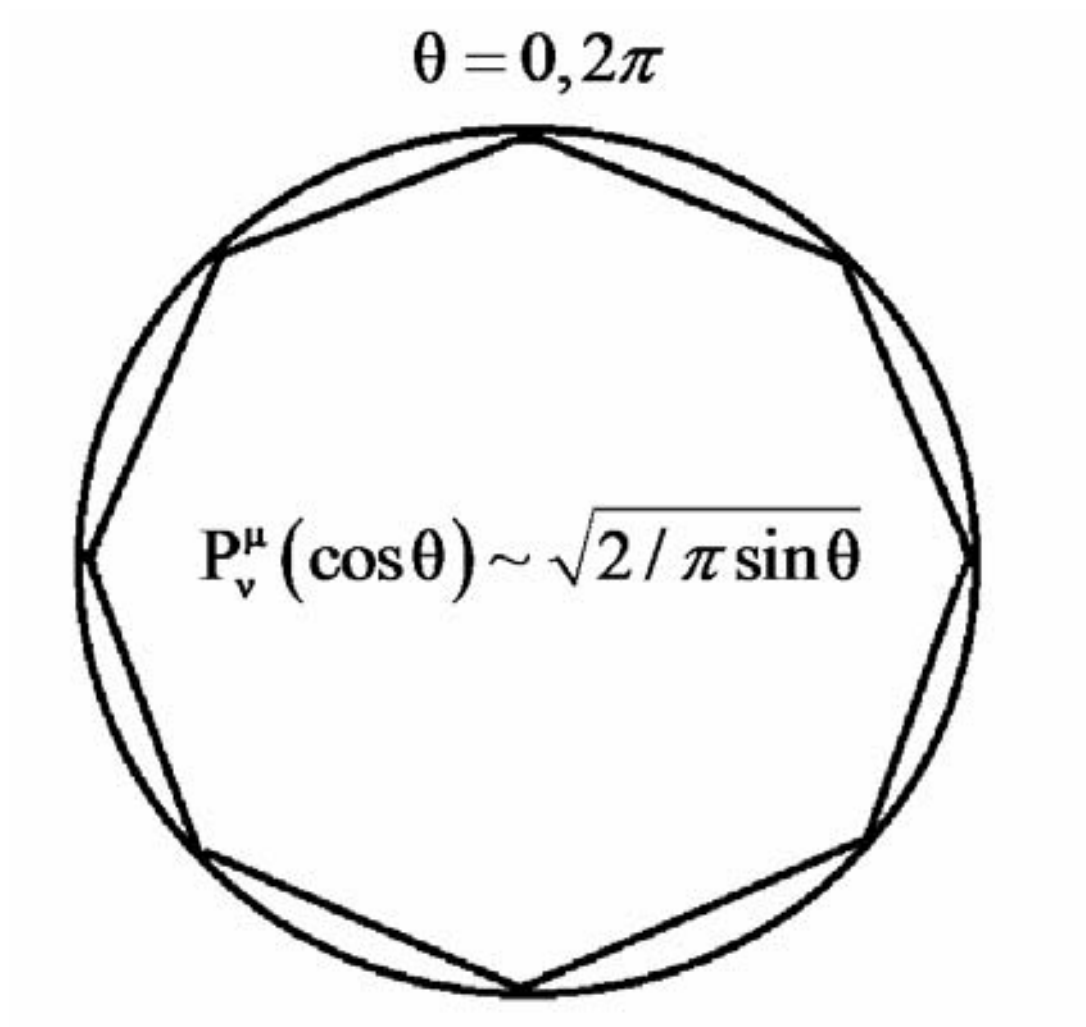

Fig. 6. Schematic pictures of the round-the-world seismic echo (see the text).

Physically, everything here is quite transparent (Fig. 6). The surface waves are excited by the mainshock, and are returned to the epicenter after revolution around the Earth. It was they induce a strong aftershock. The body waves can also create a round-the-world echo (see the broken line in Fig. 6 which illustrates the resonant ray of whispering gallery type). It is important to understand that the epicenter is the focus (caustic). The amplitude of echo increases sharply with the approach to the epicenter. The round-the-world echo is capable of inducing a strong aftershock, since the crust in the vicinity of the epicenter is in the stress-strain state for a long time after the mainshock.

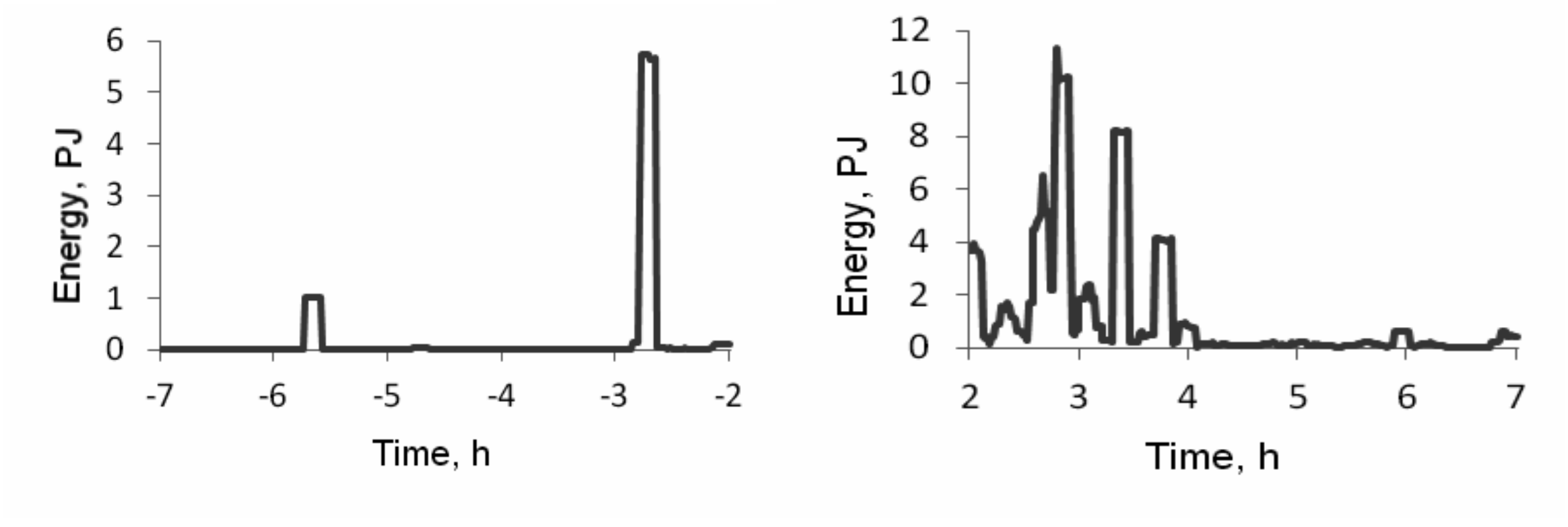

Fig. 7. The energy of foreshocks (left panel) and aftershocks (right panel) in the epicentral zones of strong earthquakes (M ≥ 7.5).

Fig. 7 on the right shows the energy of aftershocks as a function of time. As above, we used the superposed epoch analysis. The occurrence times for the earthquakes with M > 7.5 are used as the references for synchronizing the aftershocks. The peak of energy release falls on 2.8

h. The energy at the maximum is about 11 petajoules. We see the powerful cumulative effect of the converging surface waves excited by a main shock.

Let us compare the energy of aftershocks with the energy of foreshocks. Fig. 7 on the left shows the energy distribution for the foreshocks. We are seeing the powerful energy release about 3 hours before the mainshock, to be exact in -2.8 h. Let us pay attention to the symmetrical arrangement of the peaks relative to the bench mark. This is a highly interesting. Naturally, the question arises: Are not the echo-signals from foreshocks, which form a peak at -2.8 h, the triggers of mainshocks just like the echo-signals from mainshocks are the triggers of aftershocks, which form peak at +2.8 h? It is possible that at least part of the strong earthquakes is induced in this manner.

## 6. Discussion

### 6.1 On the scaling

Let us discuss the question of the similarity between the Fig. 4 (see above) and Fig. 13, which presented in the paper [Sobolev, 2011] without any reference to the catastrophe theory. According to Fig. 13, the oscillation frequency of seismic activity is reduced by about one octave for several years preceding the strong earthquake. In this case, the oscillation frequency is of the order of a few tens of nanohertz. In contrast, in our work we deal with fluctuations in the range of millihertz (see Fig. 4). Thus, the frequencies differ by four orders of magnitude in our case and in the case considered in the paper [Sobolev, 2011].

The fact that there is a decrease in the frequency of about one octave in both cases would not be worth even a mention, if not a mysterious thing. Namely, the time intervals over which frequency drop occurs at one octave are differ by four orders of magnitude, similarly as vibration frequencies are differ by four orders in these two cases. Not whether we are dealing with the scaling law? It would be highly interesting. We cannot yet answer the question, but it deserves further study.

### 6.2 The foreshock as a trigger of some aftershock

The exogenous triggers can be natural or artificial, pulsed or periodic, as well as forced or thermal, electromagnetic or mechanical; they can be terrestrial or cosmic origin. One of the priorities is to systematize and further study of triggers that affect on metastable domains of the earth's crust. We believe that the search for modulation effects of seismicity under the action of periodic triggers is important and promising. The long-term study revealed signs of the strictly periodic permanent influence of the technosphere on the seismic activity [Zotov, 2007; Guglielmi, Zotov, 2012]. The modulation of seismicity under the action of spheroidal oscillations of the Earth is also strictly periodic [Guglielmi, Zotov, 2013; Guglielmi et al., 2013]. In this case, the influence can also be regarded as permanent.

In contrast, the impact of round-the-world echo on the earthquakes is neither a permanent nor a strictly periodic. Apparently, the echo delay time varies within 2.5 - 3 h depending on the frequency and type of the seismic waves which are forming the echo. It was noted that the echo of the mainshock leads to an activation of aftershocks [Guglielmi et al., 2013]. The effect is manifested in 2.5 - 3 h after the mainshock. Now, let's remember that the activity of foreshocks increases sharply just before the mainshock (see Fig. 2). We want to pose the following question. Does not affect whether the increased activity of foreshocks on the activity of aftershocks through the mechanism of the echo? If so, then we could be understand partly the high activity of aftershocks with maximum at about 0.5 h after the mainshock.

## 7. Summary

The most important results of our work are follows. First of all, we found a sharp increase of foreshock activity 3 h before the mainshock. This property is reminiscent of the well-known activation of the ULF magnetic precursors of earthquakes. Then we observed the lowering of the characteristic frequency of foreshock sequence a few hours before the mainshock, which is consistent with a prediction of the catastrophe theory. Finally, our analysis testifies that the round-the-world seismic echo may be the trigger of mainshock. The idea is that the surface waves propagating outwards from the foreshock return back to the vicinity of the epicenter after having made a complete revolution around the Earth and induce there the mainshock.

*Acknowledgments*. We would like to acknowledge helpful conversations with Dr. A.D. Zavyalov, and Dr. O.D. Zotov. We are also thankful for valuable comments from Dr. B.I. Klain. We thank the staff of ISC for providing the catalogues of earthquakes. The work was supported by the RFBR (13-05-00066), and Programs of the Presidium RAS of Basic Research № 4 (Project 6.2).